\documentclass[a4paper, aps, prd,twocolumn,groupedaddress, showkeys, showpacs]{revtex4}
\newcommand{\haak}[1]{\left( #1 \right)}
\newcommand{\brak}[1]{\left[ #1 \right]}
\newcommand{\afg}[1]{\frac{\partial #1}{\partial t}}
\newcommand{\afgz}[1]{\frac{\partial #1}{\partial z}}
\newcommand{\ve}{\bm{v}^1}
\newcommand{\be}{\bm{B}^1}
\newcommand{\je}{\bm{j}^1}
\newcommand{\us}{c_\mathrm{s}}

\usepackage{bm, amssymb, amsfonts, amsmath, natbib, aas_macros}

\usepackage[english]{babel}
\usepackage{graphicx}

\begin{document}
\title{Gravitational waves in magnetized relativistic plasmas}
\author{J. Moortgat}
\email{moortgat@astro.kun.nl}
\homepage{http://moortgat.astro.kun.nl}
\author{J. Kuijpers}
\email{kuijpers@astro.kun.nl}
\affiliation{Department of Astrophysics, University of Nijmegen, PO Box 9010, 6500 GL Nijmegen, The Netherlands}
\date{\today}

\begin{abstract}
We study the propagation of gravitational waves ({\sc gw}) in a uniformly magnetized plasma at arbitrary angles to the magnetic field. No a priori assumptions are made about the temperature, and we consider both a plasma at rest and a plasma flowing out at ultra-relativistic velocities.
In the $3$+$1$ orthonormal tetrad description, we find that all three fundamental low-frequency plasma wave modes are excited by the {\sc gw}.
Alfv{\'e}n waves are excited by a $\times$ polarized {\sc gw}, whereas the slow and fast magneto-acoustic modes couple to the $+$ polarization. 
The slow mode, however, doesn't interact coherently with the {\sc gw}. The most relevant wave mode is the fast magneto-acoustic mode which in a strongly magnetized plasma has a vanishingly small phase lag with respect to the {\sc gw} allowing for coherent interaction over large length scales. When the background magnetic field is almost, but not entirely, parallel to the {\sc gw}'s direction of propagation even the Alfv{\'e}n waves grow to first order in the {\sc gw} amplitude.
Finally, we calculate the growth of the magneto-acoustic waves and the damping of the {\sc gw}.
\end{abstract}
\maketitle

\section{Introduction}\label{sec::intro}
The interaction of gravitational waves ({\sc gw}) with a magnetized plasma relies on the non-isotropy of the magnetic field and the fact that both {\sc gw} and electromagnetic waves propagate at the speed of light in vacuum. In vacuum, the {\sc gw} excites an electromagnetic wave propagating in the same direction as the {\sc gw} at the speed of light as was first realized by \cite{gertsenshtein}. 
Propagation through a perfect collisionless fluid does not affect a {\sc gw} \cite{gayer, kip}, but in an astrophysical plasma the electromagnetic field is coupled to the matter and, indirectly, the {\sc gw} interacts with both the electromagnetic fields and quantities such as density, pressure and currents \cite{macedo}.

For a summary of previous work on the interaction of a {\sc gw} with plasma waves we refer to the introduction of \cite{moortgat}, referred to from hereon as Paper I, and references therein and \cite{servinthesis} for more recent work.

In this paper we study the full problem of a plane fronted, monochromatic {\sc gw} of either $+$ or $\times$ polarization propagating obliquely through a  magnetized collisionless plasma. Furthermore, we don't specify whether the plasma is Poynting flux or matter dominated. Also, we allow for relativistic velocities as we want to apply our results to an ultra-relativistic force-free wind or jet. 

The outline of this paper is as follows.
We start in Sect.~\ref{sec::einstmaxw} with a discussion of the Einstein field equations ({\sc efe}) describing both the background curvature due to the static energy content and the dynamical interaction of space-time with a time dependent energy-momentum density. In the geometric optics limit (Sect.~\ref{sec::optics}) the {\sc gw} can be treated just like photons traveling on null geodesics.
In Sect.s~\ref{sec::split} -- \ref{sec::tetrad} we recapitulate how the proper reference frame of an observer (the $3+1$ orthonormal tetrad frame) is defined by taking snapshots of four-dimensional space-time 
and using ordinary vector calculus on these spatial hypersurfaces. Also, the covariant derivatives and connection coefficients for such a non-coordinate frame are summarized.
A closed set of linearized general relativistic magneto-hydrodynamic ({\sc mhd}) equations is derived in Sect.~\ref{sec::grm} and solved algebraically in Sect.~\ref{sec::waves}. We find that all the fundamental plasma waves are excited: a $+$ polarized {\sc gw} excites fast magneto-acoustic waves, which was expected from previous idealized calculations in \cite{papadopoulos2} and Paper I. In this paper we extend this result to a realistic fast mode with both electromagnetic and gas properties and find that also the slow magneto-acoustic mode is excited (Sect.~\ref{sec::msw}). A completely new result is that a $\times$ polarized {\sc gw} propagating at an angle to an ambient magnetic field  excites Alfv{\'e}n waves in the plasma  (Sect.~\ref{sec::alfven}). The space-time solutions in the comoving frame are derived in Sect.~\ref{sec::spacetime} and boosted to an ultra-relativistic wind in the the observer frame in Sect.~\ref{sec::relativistic}. In the limit where the phase velocities of the Alfv{\'e}n and fast mode approach the speed of light the waves can interact coherently, which results in linear growth of the amplitudes. For the Alfv{\'e}n mode this also requires that the angle with the magnetic field is very small (Sect.~\ref{sec::growth}).

Finally, we investigate the back-reaction on the {\sc gw} in Sect.~\ref{sec::damping} and find that as the plasma waves grow, the {\sc gw} is damped with a group velocity that decreases linearly with distance. An intuitive interpretation of some of these results is presented in Sect.~\ref{sec::interpretation}, and we end with conclusions in Sect.~\ref{sec::conclusions}.

Throughout this paper Gaussian geometrized units are adopted ($G = c = 1$). Latin indices $a \ldots e$ stand for $0, 1, 2, 3$, and $i,j,k$
for spatial components $1, 2, 3$. 

\section{Einstein-Maxwell equations}\label{sec::einstmaxw}
In Lorentz gauge, the linearized Einstein field equations ({\sc efe}) for weak gravitational waves interacting with a magneto-fluid read \cite{gravitation}:
\begin{eqnarray}\label{eq::efe}
G^{ab} \simeq -\frac{1}{2} \Box h^{ab} &=& 8 \pi \delta T^{ab},
\end{eqnarray}
where $\delta T^{ab}$ is the oscillatory part of the energy-momentum tensor. 
If we further specify to the transverse traceless ({\sc tt}) gauge and consider a {\sc gw} propagating in the $z$ direction, the only independent components of $h^{ab}$ are $h^{xx} = - h^{yy} = h_+$ and $h^{xy} = h^{yx} = h_\times$ indicating the $+$ and $\times$ {\sc gw} polarizations respectively. The individual propagation equations for the two polarizations are:
\begin{subequations}
\label{eq::gwgrowth}
\begin{eqnarray}
\Box h_+ &=& -8 \pi (\delta T^{xx} - \delta T^{yy}), \\
\Box h_\times &=& -8 \pi (\delta T^{xy} + \delta T^{yx}).
\end{eqnarray}
\end{subequations}
As an example, a {\sc gw} propagating perpendicularly to a background magnetic field $B^{0}_x$ excites perturbations $B_x^1$ that produce an oscillating cross term in the stress-energy tensor and \eqref{eq::gwgrowth} take the form \cite{gertsenshtein}:
\begin{equation}\label{eq::damping}
\Box h_+ (z,t) = 4 B^{0}_x B^{1}_x (z,t),  \qquad \Box h_\times = 0.
\end{equation}

From Maxwell's equations (coupled to the {\sc efe} via the twice contracted Bianchi identities and consequently conservation of energy-momentum $\nabla_b T^{ab} = 0$) one finds wave equations for the plasma quantities in which the {\sc gw} appears as a source term:
\begin{eqnarray}\label{eq::EMgrowth}
\hat{\Box} B_x^{1} (z,t) &\propto& h_{+} (z,t) B^0_x,
\end{eqnarray}
where $\hat{\Box}$ is some wave propagator. \eqref{eq::gwgrowth} together with \eqref{eq::EMgrowth} 
self-consistently determine the interaction between the {\sc gw} and the plasma waves (as we will work out in more detail in Sect.~\ref{sec::damping}).

\subsection{Background curvature}\label{sec::curvature}
The exact (non-linear) Einstein field equations describe the curvature of space-time due to the presence of matter and energy. This curvature is described by the contracted Riemann tensor with an associated characteristic length scale, ${\mathcal R}$, given by the magnitude of it's largest components as:
\begin{subequations}
\begin{eqnarray}
R^{ab} &=& 8 \pi\haak{ T^{ab} - \frac{1}{2} g^{ab} T}, \\\label{eq::curvature}
\frac{1}{{\mathcal R}^2} &\sim & (B^{0})^2.
\end{eqnarray}
\end{subequations}
When the interaction with a magnetic field excites electromagnetic waves growing linearly with distance, viz $B^{1} \propto B^{0}  h_+ k z$ we can see from our rough estimate \eqref{eq::curvature} that the fraction of {\sc gw} energy that is converted into electromagnetic waves is proportional to $B^{1}/h_+ \propto z/\mathcal R$. As an example, a background magnetic field comparable to the surface field of a neutron star ($B^{0}  \sim 10^{8}$~T) curves space on a scale of $\sim 10^{10}$~m. If this field would remain constant with distance, all the {\sc gw} energy would be converted to {\sc em} energy on a length scale of order ${\mathcal R}$. In reality, we will be studying a force-free plasma wind in which the magnetic field falls off linearly with distance. In this case the background curvature also decreases and the interaction length scale 
is much smaller than the radius of curvature.

The opposite limit of a weak primordial magnetic field with an extremely large spatial extent was discussed by \cite{zeldovich} who argued that it is very difficult to keep the interaction coherent on such a scale in a universe that is not a perfect vacuum.

\subsection{Geometric optics}\label{sec::optics}
The gravitational waves are assumed to be of the form:
$h_{+,\times} \sim {\mathcal H} (z) \mathrm{e}^{i\omega(z-t)}$ with a slowly varying amplitude such that $\omega {\mathcal H}(z) \gg \frac{\partial}{\partial z} {\mathcal H}(z)$ and $\frac{\lambda}{{\mathcal R}} \ll 1$. This is the short wavelength, geometric optics limit where {\sc gw}s behave as rays moving on null geodesics $k^a k_a=0$ that experience dispersion, refraction, lensing etc. The {\sc gw} move in an essentially flat Minkowski background, $\eta^{ab}$, so the full metric is $g^{ab} = \eta^{ab} + h^{ab}$ with $| h^{ab} |\ll 1$ and self-interactions (of order $h^2$) are negligible.  

We will study the propagation of transverse traceless gravitational waves in a magnetized plasma at arbitrary angle $\theta$ to a background magnetic field $\bm{B}_0$. Also we don't specify the temperature (or equivalently the pressure) regime except that the temperature equilibration time between the different particle species is short as compared to other characteristic timescales to comply with a hydrodynamic description. 
The only other non-vanishing plasma quantities in the equilibrium state are then the energy density $\mu^{0}$ and pressure $p^{0}$.
The passing {\sc gw} will excite small (first-order) perturbations in all plasma quantities, denoted as for instance: $\mu = \mu^{0} + \mu^1$.

\subsection{No coupling to unmagnetized plasma}\label{sec::nocoupling}
The stress-energy tensor for a homogeneous perfect fluid in the rest frame of an observer ($u^a = (1,0,0,0)$) is:
\begin{equation}
T^{ab} = (\rho + p) u^a u^b + p g^{ab}.
\end{equation}
A linearly polarized {\sc gw} will produce perturbations in $\rho$, $p$ and $u^i$ of order $h_{+,\times}$. However, the $\delta T^{ij}$ components are all higher order except for the trace $\delta T^{ii}$, which is purely gauge. Explicitely, in \eqref{eq::gwgrowth} $T^{xy}= T^{yx}=0$ and $T^{xx}-T^{yy} =0$.
 
Therefore a gravitational wave cannot couple to an unmagnetized perfect fluid in linearized theory \cite{kip}. Only relativistically gyrating particles can interact with a $\times$ polarized {\sc gw} through non-linear $(\rho\! +\! p) v_i v_j$ terms \cite{papadopoulosnonlinear, servin6}.

\subsection{Space-time split}\label{sec::split}
To simplify equations, we will use the so-called $3$+$1$ split of space-time \cite{ellisbook, mcdonald}.
A time-like observer moving with $4$-velocity $u^a$ perceives space as the $3$ dimensional hypersurface 
\footnote{%
The rest-space volume element is related to the $4$D element ($\epsilon^{abcd} \!= \!\epsilon^{[abcd]}$; $\epsilon^{0123}\!=\!\sqrt{|\det g|}$) by $\epsilon^{abc}\! \equiv\! \epsilon^{abcd} u_d$.} 
orthogonal to $u^a$, and $u^a$ itself as the time axis.
We can define $u^a (x^b)$ at each point in space-time as the direction of time, and define space as the {\em snapshots} of space at constant time. Subsequently, we can split equations into their space and time components by using the parallel and orthogonal projection operators $U^a_{\ b} \equiv - u^a u_b$ and $H_{ab} \equiv g_{ab} + u_a u_b$, respectively, with $U^a_{\ b} + H^a_{\ b} = \delta^a_{\ b}$.

As an example, the covariant electromagnetic Faraday tensor $F^{ab}$ (and it's dual ${\mathcal F}^{ab}  \equiv \frac{1}{2} \epsilon^{abcd} F_{cd}$) can be simplified by splitting it into its time and space components:
\begin{eqnarray}\nonumber
F^{ab} &=&  (U^a_{\ c} U^b_{\ d} + H^a_{\ c} H^b_{\ d})  F^{cd}\\\label{eq::split}
&=& u^a E^b - E^a u^b + \epsilon^{abc} B_c = \epsilon^{abc} B_c,
\end{eqnarray}
where we have defined $B_a \equiv \frac{1}{2} \epsilon_{abc} F^{bc}$ and $E^a \equiv F^{ab} u_b$ that reduce to the magnetic and electric field, respectively, in the rest frame of an observer. The last equality in \eqref{eq::split} is only valid in the ideal {\sc mhd} approximation $E^a = 0$. Since $E^a = \epsilon^{abc} u_b B_c$ or $\bm{E} = - \bm{v}\times\bm{B}$ the requirement that the comoving electric field vanishes replaces Ohm's law.
Similarly, the covariant Maxwell equations can be split into their space and time components. For a derivation we refer to \cite{ellisbook}.

\subsection{Proper reference frame}\label{sec::tetrad}
In describing the interaction of a {\sc gw} with a plasma, one has two choices for the reference frame. One is the transverse-traceless coordinate frame, discussed in the previous sections, which is tuned to a {\sc gw} with metric:
 \begin{equation}\label{eq::ttmetric}
 g_\mathrm{TT}^{ab} = \left(\begin{array}{cccc}
 -1&0&0&0\\
 0&1+h_+ (z, t)& h_\times (z, t)&0\\
 0 &h_\times (z, t)&1-h_+ (z, t)&0\\
 0&0&0&1
 \end{array}\right).
 \end{equation}
The natural reference frame of an observer, however, is an orthonormal tetrad frame ({\sc onf}).
The basis vectors that remain orthogonal in the presence of a {\sc tt} plane polarized {\sc gw} are:
\begin{equation}
\begin{array}{lcl}
 {\bm e}_0 &= & (\frac{\partial}{\partial_t}, 0, 0, 0), \\[.25cm]
  {\bm e}_1 &=& (0, (1-\frac{h_+}{2})\frac{\partial}{\partial_x}, -\frac{h_\times}{2}\frac{\partial}{\partial_y}, 0),\\[.25cm]
{\bm e}_2 &= & (0, -\frac{h_\times}{2}\frac{\partial}{\partial_x}, (1+\frac{h_+}{2})\frac{\partial}{\partial_y}, 0), \\[.25cm]
 {\bm e}_3 &=& (0, 0, 0, \frac{\partial}{\partial_z}),
\end{array}
\end{equation}
where the partial derivatives reflect the notion that in curved space-time, tangent vectors and partial derivatives are equivalent. With respect to these basis vectors the metric is $\eta^{ab}$.

Covariant derivatives are defined as: $\nabla_a T_{bc} = \bm{e}_a T_{bc} - \Gamma^d\!_{ba} T_{dc} - \Gamma^d\!_{ca} T_{bd}$, where the connection coefficients are linear combinations of the commutation functions $\gamma^a\!_{bc}$ and not derivatives of the metric as in a coordinate frame \cite{gravitation}\footnote{However, the commutation functions are defined in such a way as to give the same permutation of indices in the connection coefficients.}:
\begin{equation}
\Gamma_{abc} = \frac{1}{2} (\eta_{ad} \gamma^d\!_{cb} - \eta_{bd} \gamma^d\!_{ca} + \eta_{cd} \gamma^d\!_{ab}),
\end{equation}
and are skew in the first two indices, $\Gamma_{(ab)c}=0$. 

There seems to be some inconsistency in the literature on the explicit form of the connection coefficients.  In a coordinate frame, the connection coefficients are given by the Christoffel symbols. For the metric \eqref{eq::ttmetric} the Christoffel symbols are $\propto \dot{h}/2$ and have $12$ non-vanishing components for each polarization. 
The explicit form of these components are given in \cite{servin1} and \cite{papadopoulos2}, but it should be noted that in \cite{servin1} one of course also has the symmetric components, and in \cite{papadopoulos2} $\Gamma^0_{\ 10}$, $\Gamma^0_{\ 01}$ should be $\Gamma^1_{\ 10}$, $\Gamma^1_{\ 01}$.

In the orthonormal tetrad frame the connection coefficients only have $8$ non-vanishing components for each polarization:
\[
 \Gamma_{[02]2} = -\Gamma_{[01]1} =\frac{1}{2} \afg{h_+}\qquad
 \Gamma_{[32]2} = -\Gamma_{[31]1} =\frac{1}{2} \afgz{h_+}
\]
\[
 \Gamma_{[20]1} = -\Gamma_{[01]2} =\frac{1}{2} \afg{h_\times}\qquad
 \Gamma_{[13]2} = -\Gamma_{[32]1} =\frac{1}{2} \afgz{h_\times}
\]
where $\Gamma_{[01]1}$ stands for $\Gamma_{011}=-\Gamma_{101}$ etc.

The Einstein field equations and the equations of geodesic deviation are derived from the Riemann curvature tensor, that in the {\sc onf} is given to first order in $h_{+,\times}$ by:
\begin{equation}\label{eq::riemann}
R^a_{\ bcd} = \bm{e}_c \Gamma^{a}_{\ bd} - \bm{e}_d \Gamma^{a}_{\ bc} + {\mathcal O}[h^2].
\end{equation}
The Ricci tensor $R_{ab}$ is just a contraction of \eqref{eq::riemann} and to first order reduces to the same form as in the {\sc tt} coordinate frame. This means that the Einstein field equations in the proper reference frame also have the same form as in \eqref{eq::efe} (we will use this in 
Sect.~\ref{sec::damping} when discussing the damping of the {\sc gw}).

The driving force of a {\sc gw} on a test particle is also described through the Riemann tensor in the form of the equations of geodesic deviation:
\begin{equation}\label{eq::acceleration}
\frac{\mathrm{d}^2 x^i}{\mathrm{d} t^2} = - R_{i0j0} x^j = 
\frac{1}{2} \left(\begin{array}{ccc}\ddot{h}_+ & \ddot{h}_\times & 0\\\ddot{
h}_\times & -\ddot{h}_+ & 0 \\ 0&0&0\end{array} \right) \left(\begin{array}{c}x\\ y\\ z\end{array} \right),
\end{equation}
where $\ddot{h}_{+,\times} = \frac{\partial^2}{\partial t^2} h_{+,\times}$. 
This equation will be important for our interpretation of the interaction with a magnetic field in Sect.~\ref{sec::interpretation}.

\section{General relativistic MHD}\label{sec::grm}
Throughout this paper we assume that the {\sc mhd} approximation is valid, 
viz that the plasma is a collisionless one-fluid with negligible viscosity, resistivity 
and heat flow. In contrast to some definitions of the {\sc mhd} approximation, 
however, we do allow for relativistic velocities which means that displacement 
currents cannot be neglected and a generalized definition of the Alfv{\'en} 
velocity is needed (Sect.~\ref{sec::waves}). 
We don't restrict the angle between the {\sc gw} propagation and the background magnetic field, but without loss of generality choose it to lie in the $x$-$z$ plane.

\subsection{Coupling to the electromagnetic field}\label{sec::coupling}
A gravitational wave propagating through a uniform magnetic field $\bm{B}^0$ produces a  
Lorenz force $\bm{F}_\mathrm{L}^1$ given by the projection of Amp{\`e}re's law, $\nabla_b F^{ab} = j^a$, perpendicular to $\bm{B}^0$:
\begin{eqnarray}\nonumber
\bm{F}_\mathrm{L}^1 &=& 
\bm{j}^1\times\bm{B}^0
=
\frac{(\bm{B}^0 \cdot \nabla)\bm{B}^1}{4 \pi} - \nabla\haak{\frac{\bm{B}^0 \cdot \bm{B}^1}{4 \pi}} \\\label{eq::lorentz}
&-& \frac{|\bm{B}^0|^2}{4\pi} \afg{\ve} +  \frac{\bm{B}^0}{4\pi} \frac{\partial}{\partial t}(\ve\cdot \bm{B}^0) - \frac{\bm{j}_\mathrm{E} \times \bm{B}^0}{4\pi}.
\end{eqnarray}
Faraday's law, $\nabla_b {\mathcal F}^{ab} = 0$, is given in the $3+1$ split by:
\begin{equation}\label{eq::faraday}
\afg{\be} = (\bm{B}^0 \cdot \nabla)\bm{v}^1 - \bm{B}^0 (\nabla \cdot \ve) - \bm{j}_\mathrm{B}. 
\end{equation}
The projection of \eqref{eq::faraday} onto $\bm{B}_0$ governs the evolution of the magnetic energy density:
\begin{equation}\label{eq::faradaycontr}
\frac{\partial}{\partial t}\frac{\bm{B}^0 \cdot \bm{B}^1}{4 \pi} = 
\bm{B}^0 \cdot \nabla \frac{\ve \cdot \bm{B}^0}{4\pi}
- \frac{|\bm{B}^0|^2}{4\pi} \nabla \cdot \ve -\frac{\bm{j}_\mathrm{B}\cdot\bm{B}^0}{4\pi}.
\end{equation}
The {\sc gw} source terms in (\ref{eq::lorentz}--\ref{eq::faradaycontr}) are given by:
\begin{equation}\label{eq::jejb}
\bm{j}_\mathrm{E} = \frac{B_x^0}{2} \frac{\partial}{\partial z} \haak{\begin{array}{c}-h_\times\\h_+\\0\end{array}},
\qquad
 \bm{j}_\mathrm{B} = -\frac{B_x^0}{2} \frac{\partial}{\partial t} \haak{\begin{array}{c}h_+\\h_\times\\0\end{array}}.
\end{equation}

\subsection{Eq. of state \& Conservation of number density}\label{sec::nrdensity}
From the first law of thermodynamics, $d U = d Q - p d V$, we can find the internal energy per unit mass, $U$, as a function of the pressure, $p$, the specific volume per unit mass, $V=1/\rho$, and the heat flow that, however, vanishes under the {\sc mhd} condition: $d Q =0$. We assume an adiabatic {\em equation of state} $p = K \rho^\gamma$, where $\gamma$ is the adiabatic index, which lies in $4/3 \leq \gamma \leq 5/3$.
The total relativistic matter energy density with respect to the $4$-velocity of an ideal fluid (with the velocity of light included explicitly) is given by:
\begin{equation}\label{eq::eos}
\mu = \rho (c^2 + U) = \rho c^2 + \frac{p}{\gamma -1},
\end{equation}
and the relativistic enthalpy is defined by $w^0 = \mu^0 + p^0$. From these expressions we can derive the proper relativistic sound velocity by considering the change in pressure with $\mu$ at constant entropy:
\begin{equation}\label{eq::Usound}
c_\mathrm{s}^2 = 
\frac{\partial p}{\partial \mu}\bigg|_\mathrm{ad}  = 
\frac{\gamma p^0}{w^0}.
\end{equation}

The matter density $\rho$ can be solved from the covariant conservation of proper number density, $n = \rho/m_e$, which is given by $\nabla_a (n u^a) = 0$. In the orthonormal comoving frame to first order one finds:
\begin{equation}\label{eq::partcons}
\afg{\rho^1} + \rho_0 \nabla \cdot \ve = 0. 
\end{equation}

\subsection{Energy conservation}\label{sec::energy}
The conservation of energy and momentum follows from the divergence of the {\sc efe} (\eqref{eq::efe}) as:
\begin{equation}\label{eq::divT}
\nabla_b T^{ab} = \nabla_b \cdot [(\mu + p) u^a u^b + p g^{ab}] - F^{ab} j_b =0.
\end{equation} 
Contracting \eqref{eq::divT} with $U^c_{\ a}$ leads to the energy conservation equation:
\begin{equation}\label{eq::generalenergy}
\frac{\partial}{\partial t}[\Gamma^2(\mu + p) - p] + \nabla\cdot[\Gamma^2(\mu + p) \bm{v}] = 0,
\end{equation}
In the comoving frame ($\Gamma =1$) we can eliminate $\mu$ in favor of $p$ from \eqref{eq::eos} and \eqref{eq::partcons}:
\begin{equation}\label{eq::energy}
\afg{p^1} + \gamma p^0\nabla\cdot\ve=0.
\end{equation}
Note that from hereon, \eqref{eq::partcons} is redundant and can be dropped from our set of equations altogether. It can be solved separately to find $\rho^1$ from $\ve$.

In the cold non-relativistic limit when the internal energy is negligible with respect to the rest-mass energy ($p\ll\mu$), \eqref{eq::generalenergy} reduces to \eqref{eq::partcons} as in Paper I.

\subsection{Conservation of momentum}\label{sec::eom}
The equation of motion or momentum conservation in a $3+1$ split is projected out by $H^c_{\ a}$:
\begin{equation}\nonumber
\frac{\partial}{\partial t}[\Gamma^2(\mu + p) \bm{v}] + \nabla\cdot[\Gamma^2(\mu + p) \bm{v}\bm{v} + p \mathrm{I}] = \bm{j}\times\bm{B},
\end{equation}

\begin{equation}\label{eq::general_eom}
(\mu^0 + p^0)\afg{\ve} + \nabla p^1  = \je\times\bm{B}^0.
\end{equation}
The Lorentz force couples the matter to the electromagnetic fields and the {\sc gw} source terms, which is apparent from combining \eqref{eq::lorentz} and \eqref{eq::general_eom}:
\begin{eqnarray}\nonumber
w_\mathrm{tot}\afg{\ve} &=& - \nabla p^1+
\frac{(\bm{B}^0 \cdot \nabla)\bm{B}^1}{4 \pi} - \nabla\haak{\frac{\bm{B}^0 \cdot \bm{B}^1}{4 \pi}} \\\label{eq::eom}
 &+&  \frac{\bm{B}^0}{4\pi} \afg{(\ve\!\cdot\! \bm{B}^0)} - \frac{\bm{j}_\mathrm{E} \times \bm{B}^0}{4\pi},
\end{eqnarray}
where we have defined:
\begin{equation}\label{eq::energytot}
w_\mathrm{tot}=w^0 + \frac{|\bm{B}^0|^2}{4\pi}. 
\end{equation}

This form of the equations of motion is also found by explicitly evaluating the divergence of the electromagnetic part of the stress-energy tensor, $\nabla_b T^{ab}_\mathrm{EM} =  \frac{1}{4\pi} \nabla_b  (F^{a}_{\ c} F^{bc}\! -\! \frac{1}{4} g^{ab}  F^{cd} F_{cd}) $, instead of using Amp{\`e}re's law. The time projection, $\nabla_b T^{0b}$, then results in a conservation equation for the total energy density, e.g. \eqref{eq::faradaycontr} and \eqref{eq::energy} combined.

To summarize, in this section we have obtained a closed set of partial differential equations of $z, t$ for the $16$ variables $\bm{B}, \bm{E}, \bm{j}, \tau, \bm{v}, \mu, p, \rho$ that constitute the general relativistic {\sc mhd} description of a {\sc gw} propagating through a magnetized relativistic plasma. We will first solve these equations algebraically in Fourier and Laplace space in the next section and subsequently derive the space-time solutions in Sect.~\ref{sec::spacetime}.

\section{Plasma waves}\label{sec::waves}
In a relativistic, magnetized plasma the proper sound velocity of a compressional wave is given in \eqref{eq::Usound} and we define the Alfv{\'e}n velocity of a non-compressional shear wave in the magnetic field, $u_\mathrm{A}$, and the velocity of a mixed magneto-acoustic wave ({\sc msw}), $u_\mathrm{m}$, by:
\begin{subequations}
\begin{eqnarray}\label{eq::Ualfven}
u_\mathrm{A}^2 &=&\frac{|\bm{B}^0|^2}{4\pi w_\mathrm{tot}},\\\label{eq::Umsw}
u_\mathrm{m}^2 &=& \frac{\gamma p^0}{w_\mathrm{tot}} + \frac{|\bm{B}^0|^2}{4\pi w_\mathrm{tot}}.
\end{eqnarray}
\end{subequations}
To construct a wave equation for the plasma perturbations we take a second time derivative of \eqref{eq::eom} and  eliminate $\be$ by (\ref{eq::faraday}, \ref{eq::faradaycontr}) and $p^1$ by \eqref{eq::energy}. 
In terms of the characteristic velocities (\ref{eq::Ualfven}, \ref{eq::Umsw}) we find:
\begin{eqnarray}\nonumber
\brak{\frac{\partial^2}{\partial t^2}\! -\! u_\mathrm{m}^2\nabla\nabla\cdot}\ve \! -\!
\left[\bm{u}_\mathrm{A} \frac{\partial^2}{\partial t^2} \!-\! (\bm{u}_\mathrm{A}\cdot\nabla)\nabla\right](\ve\cdot\bm{u}_\mathrm{A}) =
\\\label{eq::waveeq}
(\bm{u}_\mathrm{A}\cdot\nabla)^2\ve - \bm{u}_\mathrm{A}(\bm{u}_\mathrm{A}\cdot\nabla)\nabla\cdot\ve+ \text{{\sc gw} terms},\qquad
\end{eqnarray}
where the {\sc gw} source terms are now given by:
\[
\sqrt{\frac{w_\mathrm{tot}}{4\pi}}
\left[\nabla\haak{\bm{j}_\mathrm{B}\cdot\bm{u}_\mathrm{A}} - \frac{\partial}{\partial t}\haak{\bm{j}_\mathrm{E}\times\bm{u}_\mathrm{A}} 
-(\bm{u}_\mathrm{A}\cdot\nabla) \bm{j}_\mathrm{B}\right].
\]

\subsection{Symmetric matrix representation}\label{sec::matrix}
We are considering {\sc gw}-propagation along the $z$-axis with the wave vector $\bm{k} = (0,0,k)$ at an arbitrary angle $\theta$ to the ambient magnetic field, which is assumed to lie in the $x\!-\!z$ plane. 
To solve the system of differential equations algebraically, we Fourier transform with respect to time and Laplace transform the spatial part to allow for growing amplitudes as in Paper I.
The wave equation \eqref{eq::waveeq} can be written in a symmetric matrix representation since the non-linear general relativistic {\sc mhd} equations form a set of symmetric hyperbolic partial differential equations:
\begin{equation}\label{eq::matrix}
D \ve = \bm{J}_\mathrm{GW}^1,
\end{equation}
where $D$ is:
{\footnotesize 
\[
\left(
\begin{array}{ccc}
\omega^2(1\!-\!u_{\mathrm{A}\bot}^2)\! -\! k^2u_{\mathrm{A}\|}^2 & 0 & - (\omega^2 \!-\! k^2)u_{\mathrm{A}\|}u_{\mathrm{A}\bot} \\
0 & \omega^2\! -\! k^2u_{\mathrm{A}\|}^2& 0 \\
- (\omega^2 \!- \! k^2)u_{\mathrm{A}\|}u_{\mathrm{A}\bot} & 0 & \omega^2(1\!-\!u_{\mathrm{A}\|}^2)\! -\! k^2(u_\mathrm{m}^2\! -\! u_{\mathrm{A}\|}^2)
\end{array}\right),
\]}

\noindent
and, for $h_{+,\times} \propto \mathrm{e}^{i\omega(z-t)}$ as discussed in Sect.~\ref{sec::optics} 
(but see Sect.~\ref{sec::damping}), the {\sc gw} source terms $\bm{J}_\mathrm{GW}^1$ are:
\begin{equation}
\bm{J}_\mathrm{GW}^1 = -\frac{i \omega^2 u_{\mathrm{A}\bot}}{\omega-k} \left(\begin{array}{c}u_{\mathrm{A}\|} h_+ \\ u_{\mathrm{A}\|} h_\times \\ -u_{\mathrm{A}\bot} h_+ \end{array}\right).
\end{equation}

\subsection{Dispersion relation}\label{sec::dispersion}
A non-trivial solution for the plasma waves in \eqref{eq::matrix} requires the determinant of $D$ to vanish.
Solving for $k = k_z (\omega)$, since $\omega$ is fixed by the driving {\sc gw}, we find six solutions:
\begin{subequations}
\begin{eqnarray}\label{eq::oma}
\omega &=& \pm k_\mathrm{A} u_\mathrm{A} \cos\theta =  \pm k_\mathrm{A} u_\mathrm{A\|},\\\label{eq::omsf}
\omega
&=& \pm \frac{k_\mathrm{s,f}}{\sqrt{2}} \sqrt{(u_\mathrm{m}^2 + c_\mathrm{s}^2 u_\mathrm{A\|}^2)}\ \sqrt{1 \pm \sqrt{(1 - \sigma)}} ,
\end{eqnarray}
\end{subequations}
where we have defined the auxiliary parameter:
\begin{equation}
\sigma (\theta) \equiv \frac{4 c_\mathrm{s}^2 u_\mathrm{A\|}^2}{(u_\mathrm{m}^2 + c_\mathrm{s}^2 u_\mathrm{A\|}^2)^2}.
\end{equation}
The negative sign in \eqref{eq::omsf} refers to the the relativistic proper {\em slow magnetosonic waves} with phase velocity $u_\mathrm{s} = \omega/k_\mathrm{s}$, and the positive sign to the {\em fast magnetosonic waves} with $u_\mathrm{f} = \omega/k_\mathrm{f}$. Together with the {\em Alfv\'en waves}, we have obtained a $6\times6$ {\sc mhd} representation that in the special-relativistic limit reduces to the equations in \cite{bram}.

In a low plasma-beta ($\beta_\mathrm{pl} = 4 \pi p/B_0^2$), e.g. strongly magnetized, plasma where $u_\mathrm{A} \gg c_\mathrm{s}$, the fast mode reduces to the magneto-acoustic mode (\eqref{eq::Umsw}) slightly altered by the presence of the gas, whereas for a high plasma-beta ($c_\mathrm{s} \gg u_\mathrm{A}$) it reduces to a sound wave with velocity \eqref{eq::Usound}. For the slow mode, the behavior is essentially the other way around. 

The angular dependence is similar: for parallel propagation ($\theta =0$), $u_\mathrm{f}\!\rightarrow\!u_\mathrm{A}$ and $u_\mathrm{s}\!\rightarrow\!c_\mathrm{s}$, whereas for perpendicular propagation ($\theta= \pi/2$) which we studied in Paper I, $u_\mathrm{f}\!\rightarrow\!u_\mathrm{m}$ and $u_\mathrm{s}\!\rightarrow\! 0$.

\subsection{Coupling}\label{sec::formalcoupling}
In this section we present a formal derivation of the solutions to the inhomogeneous wave equation \eqref{eq::matrix} following \cite{melrose}.
From \eqref{eq::matrix} we have:
\begin{equation}\label{eq::matrixinv}
\ve = D^{-1} \bm{J}_\mathrm{GW}^1= \frac{\lambda_{ij} (J^1_\mathrm{GW})_{j}}{\Lambda},
\end{equation}
where $\lambda_{ij} (\omega, k)$ and $\Lambda (\omega, k)$ are the matrix of cofactors of $D$ and its determinant, respectively. These are related by: $D_{ki}\lambda_{kj}= \delta_{ij} \Lambda$.

Each solution of $\Lambda(\omega, k)=0$ can be identified with a {\em wave mode} $M$ with $\omega = \omega (k_M)$ and $\omega (-k_M)= -\omega (k_M)$.
The determinant can be factored into these wave modes:
\begin{eqnarray}\nonumber
\Lambda(\omega, k) &=& 
(\omega^2 - k^2 u_{\mathrm{A}\|}^2)(\omega^2 - k^2 u_\mathrm{f}^2)(\omega^2 - k^2 u_\mathrm{s}^2) =0
\\\label{eq::fastm}
u_{\mathrm{f, s}}^2 &=& c_\mathrm{s} u_{\mathrm{A}\|} \haak{\frac{1\pm\sqrt{1-\sigma}}{\sqrt{\sigma}}}.
\end{eqnarray}
The unit {\em polarization vector} for a wave in mode $M$, $\bm{n}_M (k)$ with $\bm{n}_M\cdot \bm{n}^\ast_M = 1$, can be constructed from $\lambda_{ij}$:
\begin{equation}
n_{M i} (k) n^\ast_{M j} (k) = \frac{\lambda_{ij}(\omega,k_M)}{\lambda_{ii} (\omega,k_M)}.
\end{equation}

\subsection{Alfv{\'e}n waves driven by $\times$ polarized {\sc gw}}\label{sec::alfven}
The wave solutions can now be evaluated in Laplace space.
The $\lambda_{yy}$ component couples to the $\times$-polarized source term and excites Alfv{\'e}n waves, viz perturbations of the magnetic field perpendicular to the background:
\begin{subequations}
\label{eq::Lalfven}
\begin{eqnarray}\label{eq::Lvy}
v^1_y (k, \omega) &=&-\frac{i}{2} \frac{h_\times \omega u_{\mathrm{A}\|}u_{\mathrm{A}\bot}}{\omega^2 - k^2 u_\mathrm{A}^2}\ \frac{\omega + k}{\omega -k},\\\label{eq::LBy}
B^1_y (k, \omega) &=&- v_y^1 (k, \omega) \frac{B^0_x}{u_{\mathrm{A}\|}u_{\mathrm{A}\bot}} \frac{\omega+k u_{\mathrm{A}\|}^2}{\omega+k},
\end{eqnarray} 
\end{subequations}
and similarly for $E^1_{x, z}$ and $j^1_{x, z}$. Note that because the Alfv{\'e}n wave propagates obliquely with respect to the background magnetic field, the electric field is no longer divergence free and consequently the Alfv{\'e}n wave is accompanied by perturbations in the charge density: $\nabla \cdot \bm{E}^1 = i k E^1_z = 4\pi \tau^1$.

\begin{figure}[h!]
\resizebox{.7\hsize}{!}{\includegraphics{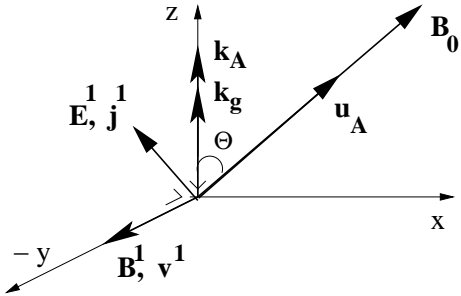}}
\caption{\label{fig::alfpolarisations}Orientation of the perturbations in the Alfv{\'e}n mode}
\end{figure}

The polarization of the Alfv{\'e}n wave components is summarized in Fig.~\ref{fig::alfpolarisations}.

\subsection{Slow and fast MSW driven by $+$ polarized {\sc gw}}\label{sec::msw}
As we expect from our considerations in Paper I, a $+$ polarized {\sc gw} excites slow and fast magneto-acoustic waves in the plasma. The velocity components are:
\begin{subequations}
\label{eq::Lv}
\begin{eqnarray}\label{eq::Lvz}
v_z^1 (k, \omega) &=& \frac{i}{2}\frac{h_+ \omega^3 u_{\mathrm{A}\bot}^2}{(\omega^2-k^2 u_\mathrm{f}^2)(\omega^2 - k^2 u_\mathrm{s}^2)}\frac{\omega + k}{\omega -k},\\\label{eq::Lvx}
v_x^1 (k,\omega) &=& -\frac{v_z (k, \omega)}{\tan\theta} \haak{1 -\frac{k^2 c_\mathrm{s}^2}{\omega^2}},
\end{eqnarray}
\end{subequations}
and from \eqref{eq::energy} one can easily find the pressure:
$
p^1 (k, \omega)  =  \frac{k}{\omega} \gamma p^0 v_z (k, \omega).
$
The magnetic component can be derived from \eqref{eq::Lv} as:
\begin{eqnarray}\label{eq::LBx}
\frac{B_x^1}{B^0} = v_z^1\sin\theta - v_x^1 \cos\theta - \frac{\omega}{\omega + k}\frac{1-u_\mathrm{A}^2}{u_\mathrm{A}^2}\frac{v_x^1}{\cos\theta}.
\end{eqnarray}
The fact that the magnetic field perturbation is orthogonal to the direction of propagation of the {\sc gw} is dictated by $\nabla\cdot\bm{B} = 0$.

\begin{figure}[h!]
\resizebox{.7\hsize}{!}{\includegraphics{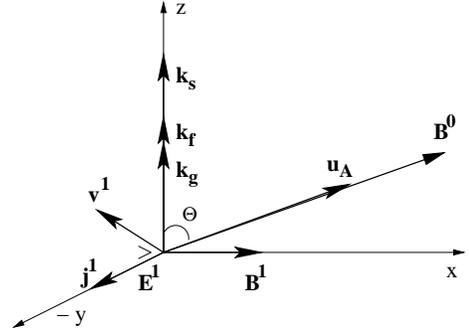}}
\caption{\label{fig::polarisations}Orientation of the perturbations in the MSW modes}
\end{figure}

The polarizations in the magneto-acoustic modes are illustrated by Fig.~\ref{fig::polarisations}.

\section{Space-time solutions}\label{sec::spacetime}
To find the solutions in space-time we apply the inverse Fourier and Laplace transformations to the results of the previous section. We define the phases of the wave modes as $\phi^\pm_\mathrm{A} = \pm k_\mathrm{A} z- \omega t$ and similarly for $\phi^\pm_\mathrm{s}$, $\phi^\pm_\mathrm{f}$, and $\phi_\mathrm{g} = \omega(z-t)$.

\subsection{Alfv{\'e}n waves}\label{sec::alfvenspace}
The most straightforward are the Alfv{\'e}n waves \eqref{eq::Lalfven} with wavenumber $k_\mathrm{A} = \omega/u_{\mathrm{A}\|}$:
\begin{equation}\label{eq::alfvenspaceB}
\frac{B_y^1}{B_x^0} \!=\!
\frac{h_\times}{4} \left[
\frac{1\!-\!u_{\mathrm{A}\|}}{1\!+\!u_{\mathrm{A}\|}}\mathrm{e}^{i \phi^-_\mathrm{A}}  \!+\! 
\frac{1\!+\!u_{\mathrm{A}\|}}{1\!-\!u_{\mathrm{A}\|}}\mathrm{e}^{ i \phi^+_\mathrm{A}}  \!-\! 
\frac{1\!+\!u_{\mathrm{A}\|}^2}{1\!-\!u_{\mathrm{A}\|}^2} 2\mathrm{e}^{i \phi_\mathrm{g} }\right]
\end{equation}
and similarly for $v_y^1(z,t)$, $E_{x,z}^1 (z,t)$, $j_{x,z}^1 (z,t)$, and $\tau^1 (z,t)$ (see Appendix~\ref{app::allalfven}). Since the Alfv{\'e}n waves are non-compressional, we do not find a $p^1$ or $\mu^1$ contribution.

\subsection{Magneto-acoustic waves}\label{sec::mswspace}
Slightly more complicated is the coupled superposition of slow and fast wave modes, polarized in the $x\!-\!z$ plane.
\begin{eqnarray}\nonumber
&&v_z^1 \!=\frac{h_+}{4} \frac{u_{\mathrm{A}\bot}^2u_\mathrm{s}^2}{u_\mathrm{f}^2\! -\! u_\mathrm{s}^2}
\left[\frac{1\!+\!u_\mathrm{s}}{1\!-\!u_\mathrm{s}} \frac{\mathrm{e}^{i\phi_\mathrm{s}^+}}{u_\mathrm{s}}
-\frac{1\!-\!u_\mathrm{s}}{1\!+\!u_\mathrm{s}} \frac{\mathrm{e}^{i\phi_\mathrm{s}^-}}{u_\mathrm{s}}
-\frac{4\mathrm{e}^{i\phi_\mathrm{g}}}{1-u_\mathrm{s}^2}  \right]
 \\
&&- \frac{h_+}{4} \frac{u_{\mathrm{A}\bot}^2u_\mathrm{f}^2}{u_\mathrm{f}^2\! -\! u_\mathrm{s}^2} \left[
\frac{1\!+\!u_\mathrm{f}}{1\!-\!u_\mathrm{f}} \frac{\mathrm{e}^{i\phi_\mathrm{f}^+}}{u_\mathrm{f}}
-\frac{1\!-\!u_\mathrm{f}}{1\!+\!u_\mathrm{f}} \frac{\mathrm{e}^{i\phi_\mathrm{f}^-}}{u_\mathrm{f}}
-\frac{4 \mathrm{e}^{i\phi_\mathrm{g}}}{1-u_\mathrm{f}^2} 
\right]
\end{eqnarray}
and:
\begin{eqnarray}
&&v_x^1\tan\theta = \frac{h_+}{4} \frac{\us^2 u_{\mathrm{A}\bot}^2}{u_\mathrm{f}^2\! -\! u_\mathrm{s}^2} \left[
\frac{1\!+\!u_\mathrm{f}}{1\!-\!u_\mathrm{f}} \frac{\mathrm{e}^{i\phi_\mathrm{f}^+}}{u_\mathrm{f}}
-\frac{1\!-\!u_\mathrm{f}}{1\!+\!u_\mathrm{f}} \frac{\mathrm{e}^{i\phi_\mathrm{f}^-}}{u_\mathrm{f}}
\right. \\\nonumber
&&\left. -\frac{4u_\mathrm{f}^2 \mathrm{e}^{i\phi_\mathrm{g}}}{1-u_\mathrm{f}^2} - \frac{1\!+\!u_\mathrm{s}}{1\!-\!u_\mathrm{s}} \frac{\mathrm{e}^{i\phi_\mathrm{s}^+}}{u_\mathrm{s}}
+\frac{1\!-\!u_\mathrm{s}}{1\!+\!u_\mathrm{s}} \frac{\mathrm{e}^{i\phi_\mathrm{s}^-}}{u_\mathrm{s}}
+\frac{4u_\mathrm{s}^2\mathrm{e}^{i\phi_\mathrm{g}}}{1-u_\mathrm{s}^2}  \right] - v_z^1
\end{eqnarray}
Note that for $c_\mathrm{s} \!\downarrow \!0$, $\bm{v}^1\!\perp\! \bm{B}^0$.  If also $\theta \!=\! \pi/2$ the limiting behavior is: $u_\mathrm{f} \!\rightarrow\! u_\mathrm{A}$, $u_\mathrm{s}\! \rightarrow\!0$ and we retrieve our original idealized solution.

The remaining magneto-acoustic wave components, $B_x^1 (z,t)$, $E_y^1 (z,t)$, $j_y^1 (z,t)$, $p^1 (z,t)$, and $\mu^1 (z,t)$, are equivalent superpositions of slow and fast waves but with different relative amplitudes, summarized in Appendix~\ref{app::allmsw}.

\subsection{Growth}\label{sec::growth}
In general, for arbitrary $u_\mathrm{A},c_\mathrm{s}$, and $\theta$, we always find $u_\mathrm{f} > u_\mathrm{s}$. 
In the limit where the phase velocity of the fast mode approaches the speed of light, $u_\mathrm{f} \uparrow 1$, coherent interaction with the {\sc gw} is possible. The amplitude of the retreating fast wave ($\propto (1+u_\mathrm{f})^{-1}$) is negligible with respect to the forward wave ($\propto (1-u_\mathrm{f})^{-1}$). The forward wave grows linearly with distance because with $k_\mathrm{f} = \omega/u_\mathrm{f} = \omega + \Delta k$ to first order in $\Delta k$ we have:
\begin{equation}
\frac{\mathrm{e}^{i k_\mathrm{f} z} - \mathrm{e}^{i \omega z}}{1-u_\mathrm{f}}=
\frac{\omega}{u_\mathrm{f}\Delta k} \mathrm{e}^{i \omega z} i \Delta k\  z.
\end{equation}
This limiting behavior is the same for all magneto-acoustic components (see \eqref{eq::mswsols}). 

For the Alfv{\'e}n waves, the conditions for growing solutions are less favorable because of the angular dependence of the resonance condition. 
To interact coherently with the {\sc gw}, the phase speed of the Alfv{\'e}n wave ($v_\mathrm{ph} = u_\mathrm{A} \cos\theta$) has to approach the speed of light. On the one hand this means that the wave vector of the Alfv{\'e}n wave should be almost parallel to the background magnetic field, but on the other hand, the magnetic field should also have a transverse component, since the amplitudes are proportional to $B^0 \sin\theta \approx B^0 \theta$ for small angles.
Explicitly:
\begin{equation}\label{eq::alfvengrowth}
 \frac{B_y^1 (z,t)}{B^0}= -v_y^1 (z,t) \sim \frac{\theta h_\times}{2} \omega z\Im[\mathrm{e}^{i \phi_\mathrm{g}}] + {\mathcal O}[\theta^2].
\end{equation}

The excited slow magnetosonic wave is a purely oscillatory wave propagating both in the forward and the backward directions.

\section{A warm relativistic plasma wind}\label{sec::relativistic}
We now want to consider a warm relativistic plasma wind flowing out in the $z$ direction with constant (background) velocity $\beta$ and corresponding Lorentz factor $\gamma_\mathrm{tot} = 1/\sqrt{1-\beta^2}$.
We are allowed to use simple Lorentz transformations in this general relativistic treatment, because the whole concept of gravitational waves as small perturbations of the background space-time (linearized theory), relies on the fact that we can treat the {\sc gw}s as a field living in flat space-time as long as the scale of curvature is much larger than the wavelength of the {\sc gw}s: ${\mathcal R} \gg \lambda_\mathrm{GW}$, which was verified in Sect. \ref{sec::optics}.

From now on, all quantities derived in the previous sections for the comoving frame will be denoted by primes. Since the phase of plane waves, $\phi = k_a x^a$, is invariant under Lorentz transformations, a boost mostly implies addition of velocities.  We then have $u_f^\prime  = \frac{u_\mathrm{f} - \beta}{1-\beta u_\mathrm{f}}$, $u_s^\prime = \frac{u_\mathrm{s} - \beta}{1-\beta u_\mathrm{s}}$, and $u^\prime_{\mathrm{A}\|} = \frac{u_{\mathrm{A}\|} -\beta}{1-\beta u_{\mathrm{A}\|}}$ for the phase velocities and:
\begin{eqnarray}\label{eq::vtransform}
v^1_z \approx \frac{(v_z^{1})^\prime}{\gamma^2} &,&
v^1_{x, y} \approx\frac{(v_{x, y}^{1})^\prime}{\gamma}.
\end{eqnarray}

The boosted background magnetic field is $\bm{B}^0 = (\gamma B^{0\prime}_x, 0, B^{0\prime}_z)$ and in the laboratory frame one finds a zeroth order electric field $\bm{E}^0 = - \beta \times \bm{B}^0 = (0, - \beta\gamma B^{0\prime}_x,0)$.

\subsection{Relativistic Alfv{\'e}n waves}\label{sec::relalfven}
We define $\gamma_{u_\mathrm{A}}^2 = 1/(1- u_{\mathrm{A}\|}^2) + {\mathcal O}[\theta^2]$ as the Lorentz factor associated with the Alfv{\'e}n speed for small $\theta$ and boost \eqref{eq::alfvenspaceV} to the laboratory frame as an example:
\begin{eqnarray}\nonumber
&&v_y^1 = \frac{h_\times}{4} \frac{u_{\mathrm{A}\bot} \gamma_{u_\mathrm{A}}^2}{1-u_{\mathrm{A}\|} \beta} 
\left\{4
 [u_{\mathrm{A}\|} (1+\beta^2) - \beta(1+u_{\mathrm{A}\|}^2)] \mathrm{e}^{i\phi_\mathrm{g}}
 \right. \\\label{eq::relvy}
&&\left.
\haak{\frac{1+u_{\mathrm{A}\|}}{1+\beta}}^2
\frac{\mathrm{e}^{i \phi^+_\mathrm{A}}}{\gamma^4}
- \haak{\frac{1+\beta}{1+u_{\mathrm{A}\|}}}^2
\frac{\mathrm{e}^{i\phi^-_\mathrm{A}}}{\gamma_{u_\mathrm{A}}^4}
\right\}.
\end{eqnarray}
Or in the ultra-relativistic limit $u_\mathrm{A} \uparrow 1$ (and $\beta\simeq 1$):
\begin{equation}
v_y^1 (z,t)
\approx -\frac{h_\times u_{\mathrm{A}\bot}}{4\gamma^2}  \omega z \Im[\mathrm{e}^{i \phi_\mathrm{g}}].
\end{equation}

\subsection{Relativistic MSW}
In a Poynting flux dominated force-free plasma wind, where the magnetic energy density strongly dominates the matter density, the plasma flows out at ultra-relativistic velocities. In this regime the phase velocity of the fast mode approaches the Alfv{\'e}n velocity, $u_\mathrm{f} \simeq u_\mathrm{A} \uparrow 1$ and the phase velocity of the slow mode becomes negligible, $u_\mathrm{s}\downarrow 0$. The solutions are therefore quite similar to those found in Paper I, the main difference being the angular dependence. 
For instance:
\begin{subequations}
\begin{eqnarray}
v_x &=& \frac{v_x^\prime}{\gamma} \simeq \frac{h_+}{4 \gamma^3} \frac{\sin^2\theta}{(1-\beta\cos\theta)^2} \omega z\  \Im\left[\mathrm{e}^{i\phi_\mathrm{g}} \right],\\
v_z &=& \frac{v_z^\prime}{\gamma^2} \simeq \frac{h_+}{4 \gamma^4} \frac{\sin\theta(\cos\theta-\beta)}{(1-\beta\cos\theta)^2} \omega z\  \Im\left[\mathrm{e}^{i\phi_\mathrm{g}} \right] .
\end{eqnarray}
\end{subequations}
Appendix~\ref{app::allmswrel} lists the limiting behavior of all the remaining {\sc msw} components in a relativistic wind.

\section{Damping of the GW}\label{sec::damping}
To find an evolution equation for the gravitational waves, we project the transverse traceless part of the stress-energy perturbation as in \eqref{eq::gwgrowth} (as we discussed at the end of Sect.~\ref{sec::tetrad}, \eqref{eq::efe} and \eqref{eq::gwgrowth} also hold in the {\sc onf}).

The perturbation of the matter part only has $\delta T_\mathrm{TT}^{xx} = \delta T_\mathrm{TT}^{yy} = p^1(z,t)$ which is purely gauge (the trace can be removed by a gauge transformation). Only the magnetic field interacts directly with the gravitational waves. All the other perturbations are excited through the {\sc mhd} processes in the plasma. In the general case where oblique $h_+$ {\sc gw} excite slow and fast magneto-acoustic waves and $h_\times$ {\sc gw} excite Alfv{\'e}n waves, \eqref{eq::damping} is replaced with:
\begin{equation}\label{eq::gwevolution}
\Box h_+ =  4 B_x^0 B_x^1, \quad
 \Box h_\times = 4 B_x^0 B_y^1.
\end{equation}

As we discussed in Sect.~\ref{sec::optics} we have assumed that the {\sc gw} have a slowly varying amplitude ${\mathcal H} (z)$. We have neglected this variation in studying the interaction with the plasma since in the short-wavelength approximation all the derivatives have $\frac{\partial}{\partial z} {\mathcal H}(z)\! \ll\! \omega {\mathcal H}(z) $. It is however this slowly varying amplitude that describes the damping of the {\sc gw}. We could find an order 
of magnitude expression for ${\mathcal H} (z)$ by integrating \eqref{eq::gwevolution} using \eqref{eq::mswspaceB} and \eqref{eq::alfvenspaceB}, but this wouldn't be entirely self-consistent since in deriving \eqref{eq::mswspaceB} and \eqref{eq::alfvenspaceB} we already assumed that $\omega = k$ for the {\sc gw}.

It is, however, possible to derive a dispersion relation for the damped {\sc gw} and the excited {\sc mhd} waves simultaneously in a self-consistent way. If we assume that the {\sc gw} oscillates at a fixed frequency but leaves the spatial dependence unspecified (e.g. a boundary value problem with $h (z,t) \propto h(z) \mathrm{e}^{-i\omega t}$) we can still solve the full set of {\sc mhd} equations in Laplace space. For the 
magnetic field we find:
\begin{subequations}
\label{eq::LBxy}
\begin{eqnarray}
B^1_y (k, \omega) &=& \frac{B^0_x h_\times (k, \omega)}{2} \frac{\omega^2 + k^2 u_{\mathrm{A}\|}^2}{\omega^2 - k^2 u_{\mathrm{A}\|}^2},\\
B^1_x (k, \omega) &=& \frac{B^0_x h_+ (k, \omega)}{2} \frac{\omega^2 + k^2 u_{\mathrm{A}}^2}{\omega^2 - k^2 u_{\mathrm{A}}^2},
\end{eqnarray}
\end{subequations}
where the second expression is in the limit of a Poynting flux dominated wind with $u_\mathrm{A}\! \gg\! c_\mathrm{s}\downarrow 0$ and $h_{+, \times} (k, \omega)$ are the Laplace transforms of $h_{+, \times} (z,t)$ (the reason that \eqref{eq::LBxy} look different from \eqref{eq::LBy} and \eqref{eq::LBx} is that derivatives of $h_{+, \times}$ depend on both $\omega$ and $k$ in this more general case).
 
 If we insert \eqref{eq::LBxy} in the Laplace transform of \eqref{eq::gwevolution}, $h_{+, \times}(k, \omega) $ drop out and we find the 
 self-consistent dispersion relation for the coupled fast magnetosonic - gravitational ($h_+$) mode:
\begin{equation}\label{eq::Ldisp}
\omega^2 - k^2 = 2 (B^0_x)^2 \frac{\omega^2 + k^2 u_{\mathrm{A}}^2}{\omega^2 - k^2 u_{\mathrm{A}}^2},
\end{equation}
and the coupled Alfv{\'e}n - gravitational ($h_\times$) mode which looks the same but with $u_\mathrm{A}$ replaced with $u_{\mathrm{A}\|}$.

The two modes allowed by \eqref{eq::Ldisp} are given by:
\begin{equation}\label{eq::fulldisp}
\frac{k^2_{1,2} u_{\mathrm{A}}^2 }{1\!-\! u_{\mathrm{A}}^2} = \frac{\omega^2}{2}\left[ \epsilon + \frac{1\!+\!u_{\mathrm{A}}^2}{1\!-\!u_{\mathrm{A}}^2} 
\pm \sqrt{1+ 2 \epsilon  \frac{3\!+\!u_{\mathrm{A}}^2}{1\!-\!u_{\mathrm{A}}^2} + \epsilon^2} \right],
\end{equation}
in terms of $\epsilon = \left(\frac{B^0_x}{\omega}\right)^2\frac{2 u_{\mathrm{A}}^2}{1-u_{\mathrm{A}}^2} = 
\frac{ (B^0_x)^2}{\omega\Delta k}\frac{2 u_{\mathrm{A}}}{1+u_{\mathrm{A}}}$.
For small $\epsilon$, \eqref{eq::fulldisp} reduces to:
\begin{subequations}
\label{eq::limit}
\begin{eqnarray}\label{eq::gwmode}
k_1^2 &=& \omega^2\left(1 -  \epsilon \frac{1+ u_{\mathrm{A}}^2}{u_{\mathrm{A}}^2} \right) + \mathcal{O}[\epsilon^2], \\\label{eq::mswmode}
k_2^2 u_{\mathrm{A}}^2 &=& \omega^2 \left(1+ 2 \epsilon \right) + \mathcal{O}[\epsilon^2] .
\end{eqnarray}
\end{subequations}
Mode $1$ corresponds to a superluminal mode with phase velocity $\omega/k_1 >1$ and mode $2$ is subluminal:
\[
\begin{array}{rclcrcl}
v_{\mathrm{ph},1}^2 &=& 1 +  \frac{2(B_x^0)^2}{\omega^2} \frac{1+u_{\mathrm{A}}^2}{1-u_{\mathrm{A}}^2} &,&
v_{\mathrm{gr},1}^2 &=& 1 - \frac{2(B_x^0)^2}{\omega^2}\frac{1+u_{\mathrm{A}}^2}{1-u_{\mathrm{A}}^2}\\
\frac{v_{\mathrm{ph}, 2}^2}{u_{\mathrm{A}}^2} &= &1 - \frac{(2 B_x^0)^2}{\omega^2} \frac{u_{\mathrm{A}}^2}{1-u_{\mathrm{A}}^2} &,&
\frac{v_{\mathrm{gr}, 2}^2}{u_{\mathrm{A}}^2} & =& 1 + \frac{(2 B_x^0)^2}{\omega^2} \frac{u_{\mathrm{A}}^2}{1-u_{\mathrm{A}}^2}
\end{array}
\]
with $v_{\mathrm{ph},1} v_{\mathrm{gr},1} = 1$ and $v_{\mathrm{ph},2} v_{\mathrm{gr},2} = u_{\mathrm{A}}^2$.  Both modes have group velocity $\partial \omega/\partial k_{1,2} <1$ as long as (consistent with \cite{photonfreq}):  
\begin{equation}\label{eq::constraint}
\frac{8\pi G}{c^2}\frac{(B^0_x)^2}{\mu_0} < \omega (\Delta k) c.
\end{equation}

In the limit $\epsilon \ll 1$, \eqref{eq::mswmode} reduces to a fast magnetosonic wave in flat space-time (or, equivalently, to the Alfv{\'e}n mode for $h_\times$). Mode \eqref{eq::gwmode} reduces to a vacuum {\sc gw}, justifying our approximation $h_{+,\times} \propto \mathrm{e}^{i\omega(z-t)}$  in deriving the plasma perturbations in Sect.~\ref{sec::waves}.

\section{Interpretation}\label{sec::interpretation}
In this section we present an intuitive interpretation of our main results: that a $+$ polarized {\sc gw} excites magneto-acoustic waves and a $\times$ polarized {\sc gw} excites Alfv{\'e}n waves in a uniform magnetic field. 

The driving force exerted by a {\sc gw} on test particles is described by \eqref{eq::acceleration}.
These equations can be illustrated by force lines in the plane orthogonal to the propagation of the {\sc gw} \cite{gravitation}. 
Integrating \eqref{eq::acceleration} twice with respect to time results in the well known equations of spatial deviations of test masses in an interferometer detector such as {\sc ligo}: 
\begin{equation}\label{eq::detector}
\delta x = \frac{1}{2} (h_+ x_0 + h_\times y_0), 
\qquad
\delta y = \frac{1}{2} (h_\times x_0 - h_+ y_0).
\end{equation}
In the {\sc mhd} limit these test particles are `glued' to the magnetic field lines, so the magnetic field will exhibit the same behavior (in fact the presence of the plasma is not required, the magnetic field lines can just be viewed as parametrized by \eqref{eq::detector}). Since the action of the {\sc gw} is only in the $x-y$ plane, and the magnetic field lies in the $x-z$ plane we expect that a $+$ polarized {\sc gw} results in:
\begin{equation}\label{eq::bx}
\delta B_x \propto \frac{1}{2} h_+ B_x^0, 
\qquad 
\delta B_y \propto \frac{1}{2} h_+ B_y^0 = 0,
\end{equation}
and a $\times$ polarized {\sc gw} excites:
\begin{equation}
\delta B_x \propto \frac{1}{2} h_\times B_y^0 =0,
\qquad
 \delta B_y \propto \frac{1}{2} h_\times B_x^0.
\end{equation}
This is exactly what we found in the mathematical treatment of the previous section. Perturbations in the other directions are higher order effects.

\begin{figure}[h!]
\resizebox{.54\hsize}{!}{\includegraphics{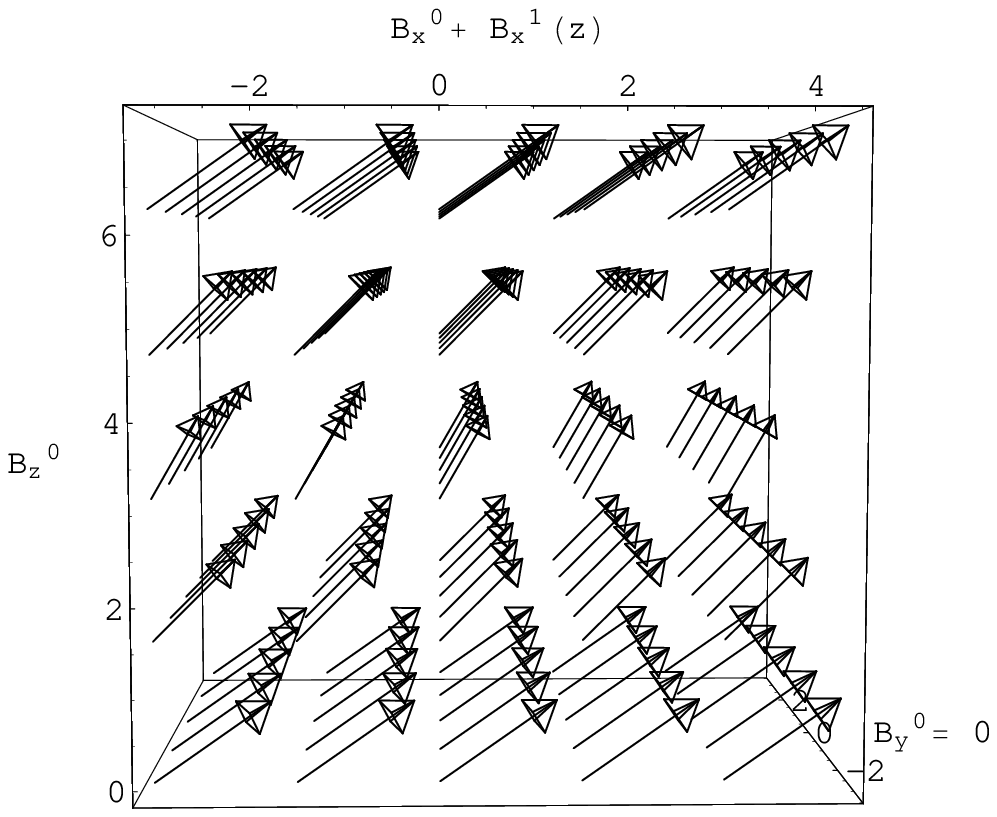}}
\resizebox{.44\hsize}{!}{\includegraphics{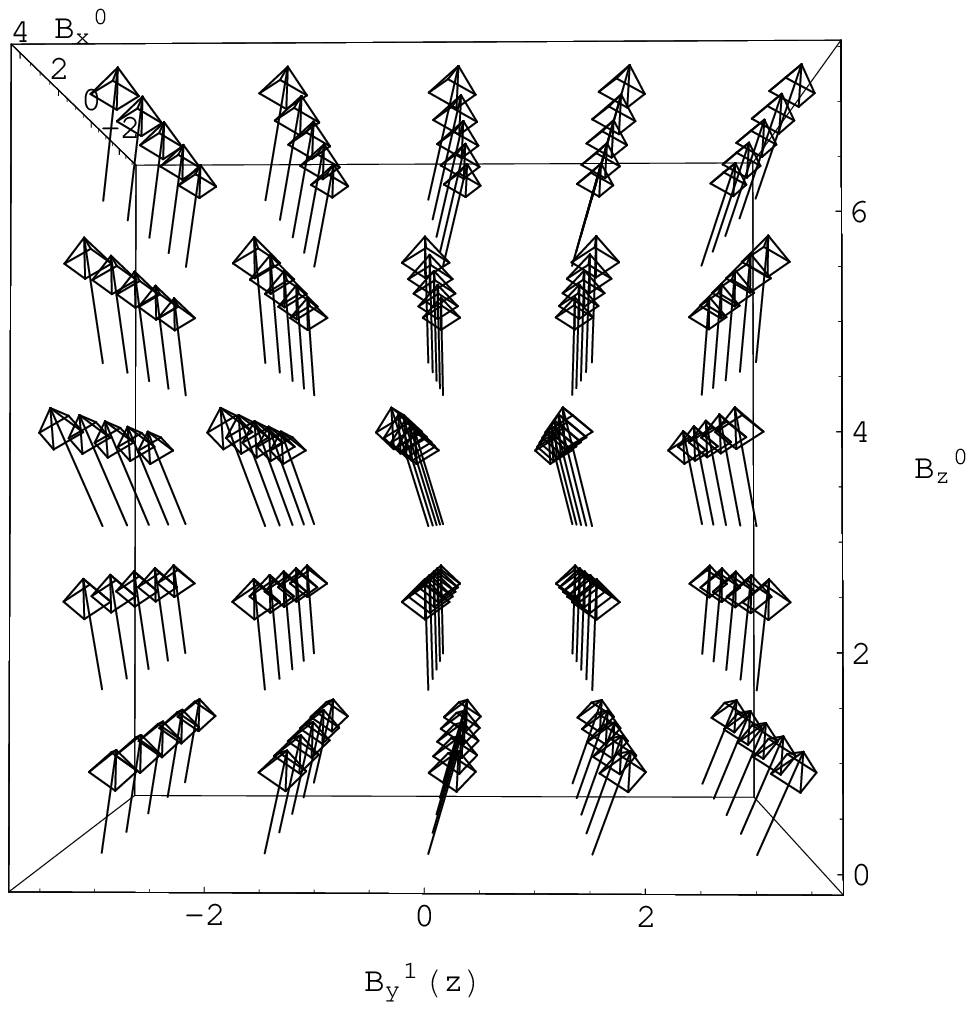}}
\caption{\label{fig::vectorfields}The {\sc msw} (left) and Alfv{\'e}n mode (right) illustrated as an oscillating vector field. The $x-y$ axes in the right figure are rotated by $\pi/2 $ with respect to those in the left figure.}
\end{figure}
Fig.~\ref{fig::vectorfields} gives a schematic illustration of the perturbed magnetic field in the two wave modes. The left figure shows the vector field (in arbitrary units) for the oblique magneto-acoustic wave propagating in the $z$-direction. The perturbations are highly exaggerated to emphasize the effect. 
Amplification of the magnetic field occurs when $B_x^0$ is amplified and dilution when $B_x^0$ is suppressed. Since $B_z^0$ is constant the total magnetic field has an overal wavy pattern.

Alfv{\'e}n waves are non-compressional and only set up a vibration in the field lines perpendicular to the background field ($B_y^1$). This is illustrated in the right figure, where the axes are rotated to emphasize the $y-z$ plane.

In a pulsar environment the plasma initially flows out {\em along} the open field lines but develops into a force-free wind outside the light-cylinder in which the toroidal component of the magnetic field dominates the poloidal one. Here the magnetic field is predominantly perpendicular to the radial propagation of the wind as illustrated by Fig.~\ref{fig::gwbdance}.  Gravitational waves would mainly excite Alfv{\'e}n waves in the former region, whereas in the latter case the magneto-acoustic waves are favored. In the relativistic wind the {\sc gw} frequency is red-shifted and the interaction is suppressed by $\gamma^{-2}$, but the interaction length scale becomes very large.

\section{Conclusions}\label{sec::conclusions}
\begin{figure}
\resizebox{.94\hsize}{!}{\includegraphics{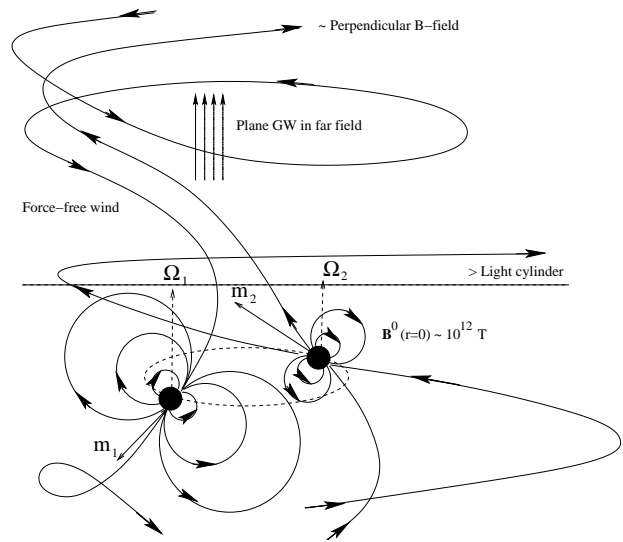}}
\caption{\label{fig::gwbdance}Merging neutron star binary}
\end{figure}
We have studied the propagation properties of a plane polarized gravitational wave in a magnetized astrophysical plasma in the most general case. Both polarizations of the {\sc gw} have been taken into account. Oblique propagation with respect to the background magnetic field was studied including pressure terms, and relativistic velocities. The only approximations in this treatment are the {\sc mhd} conditions for the plasma, the linear perturbative approach, the geometric optics or small wavelength limit for the gravitational waves, and our assumed geometry of the interaction region.

The result is a very rich astrophysical problem, where all three fundamental plasma wave modes are excited, two of which can interact coherently with the driving gravitational waves, and as a result grow linearly with distance, dissipating {\sc gw} energy into the plasma.

Alfv{\'e}n waves are excited already in linearized theory by $\times$ polarized gravitation waves propagating at an angle with respect to an ambient magnetic field, a result that as far as we know has not been found before (note that both conditions, oblique {\sc gw} propagation and $\times$ polarization, are required). The Alfv{\'e}n waves are non-compressional shear waves that have orthogonal electromagnetic components with a corresponding drift velocity in the plasma, a current flowing along the electric field and a deviation from charge neutrality caused by the divergence of the electric field, but no pressure or density components. To interact coherently with a {\sc gw} one has to find the optimum of longest interaction length scale and significant amplitude as a function of the angle between the Alfv{\'e}n wave vector and background magnetic field as was discussed in Sect.~\ref{sec::growth}. 

As we already derived in a Paper I, $+$ polarized gravitational waves excite magneto-acoustic waves propagating parallel to the gravitational waves. In this paper we have generalized our previous treatment to include the oblique magnetic field and a relativistic equation of state with non-vanishing pressure. As a result we find both the slow and the fast magneto-acoustic waves with phase velocities that depend on both the electromagnetic and the matter properties of the plasma. 

The magneto-acoustic wave modes are compressional waves, that excite pressure, density, and magnetic field gradients along their wave vector direction, but no perturbation of charge neutrality. The electric and the magnetic field perturbations and the wave vector are mutually orthogonal ($\bm{E}^1 \bot \bm{B}^1 \bot \bm{k}_\mathrm{s,f}$), but the drift velocity is no longer exactly perpendicular to the magnetic field due to the pressure. Of course, the plasma motion in a {\sc gw} is non-compressional but it generates magnetic field compression if it propagates across a magnetic field either in a vacuum or in an ideal frozen-in plasma, and hence couples to the magneto-acoustic wave.

As to the slow {\sc msw}, no coherent interaction can occur with the {\sc gw} since the phase velocity of the slow mode is always much smaller than that of the gravitational waves. Therefore the amplitude does not grow in time or with distance and cannot become significant.

The most effective interaction occurs between the {\sc gw} and the fast magneto-acoustic wave. The phase velocity of those waves can easily approach the speed of light in a strongly magnetized plasma (in the limit $u_\mathrm{A}\! \gg\! c_\mathrm{s}$) as we derived in Paper I and therefore the waves will grow linearly with distance. As the plasma waves grow, the amplitude of the gravitational waves decays correspondingly but as long as the total length of the interaction region is smaller than the background curvature produced by the plasma itself, this will only be a small fraction of the total gravitational wave energy.

Astrophysical applications of the above interactions lie in sources of strong gravitational waves embedded in strongly magnetized plasmas. Examples are non-spherical rotating neutron stars, fast rotating neutron stars that are unstable to torsional oscillations, non-spherical supernovae collapses, magnetars, and the progenitors of gamma-ray bursts (non-spherical collapse of a massive star or a merging neutron star binary).
Most of these sources are probably accompanied by an extended strongly magnetized and force-free plasma wind, flowing out at ultra-relativistic velocities. Since the coupling constant of gravitational waves is exceedingly small, only in the most extreme sources will the interaction with the plasma be significant. 

Fig.~\ref{fig::gwbdance} shows as an example a binary neutron star as a gamma-ray burst progenitor.
For such a merging binary with a magnetar class surface magnetic field 
of $10^{12}$~T \cite{ibrahim}, an angular frequency at the end of the spiral-in phase of the order of $1$~kHz and a force-free wind flowing out with a Lorentz factor of $\Gamma \sim 100$ up to a fraction of a light year, a total energy of $10^{43}$~J can be transferred from the gravitational waves to the wind.

\begin{acknowledgements}
We would like to thank the anonymous referee for valuable comments and suggestions. This work was supported by the Dutch Research School for Astronomy, {\sc nova}.
\end{acknowledgements}

\appendix*
\section{All space-time solutions}
\subsection{Comoving Alfv{\'e}n wave components}\label{app::allalfven}
The components of the Alfv{\'e}n wave can be derived most easily by starting with it's velocity component:
\begin{equation}\label{eq::alfvenspaceV}
\frac{v_y^1}{u_{\mathrm{A}\bot}} =
\frac{h_\times}{4}\left[
\frac{1\!+\!u_{\mathrm{A}\|}}{1\!-\!u_{\mathrm{A}\|}} \mathrm{e}^{ i \phi^+_\mathrm{A} } -
\frac{1\!-\!u_{\mathrm{A}\|}}{1\!+\!u_{\mathrm{A}\|}} \mathrm{e}^{i \phi^-_\mathrm{A} } - 
\frac{4 u_{\mathrm{A}\|}\mathrm{e}^{i \phi_\mathrm{g} }}{1\!-\!u_{\mathrm{A}\|}^2}\right].
\end{equation}
From $\bm{E}^1 = - \bm{v}\times\bm{B}^0$ we find the electric field:
\[
\frac{E_x^1 (z,t)}{B^0_z} = - \frac{E_z^1 (z,t)}{B^0_x} =- v_y^1 (z,t),
\]
whereas the current density follows from Amp{\`e}re's law:
\[
j_x^1(z,t) = -\frac{i \omega}{4 \pi} \frac{1-u_{\mathrm{A}\|}^2}{u_{\mathrm{A}\|}^2}  E_x^1 (z,t),
\quad
j_z^1(z,t) =\frac{i \omega}{4 \pi} E_z^1 (z,t).
\]
The charge density is given by $\nabla\cdot\bm{E}^1 = 4 \pi \tau^1$: 
\[
\frac{\tau^1 (z,t)}{B^0_x \tan\theta} =
\frac{i \omega h_\times }{4}\left[
\frac{1\!+\!u_{\mathrm{A}\|}}{1\!-\!u_{\mathrm{A}\|}} \mathrm{e}^{ i \phi^+_\mathrm{A} } \!+\!
\frac{1\!-\!u_{\mathrm{A}\|}}{1\!+\!u_{\mathrm{A}\|}} \mathrm{e}^{i \phi^-_\mathrm{A} } \!-\! 
\frac{4 u_{\mathrm{A}\|}^2\mathrm{e}^{i \phi_\mathrm{g} }}{1\!-\!u_{\mathrm{A}\|}^2}\right].
\]

\subsection{Comoving MSW components}\label{app::allmsw}
The most complicated {\sc msw} component is the magnetic field that clearly betrays the mixed nature (gas and electromagnetic fields) of this mode. 
\begin{equation}\label{eq::mswspaceB}
B_x^1 (z,t) = \frac{B_x^0 h_+}{4(u_\mathrm{f}^2\! -\! u_\mathrm{s}^2)}[C (z,t) + D (z,t)]
\end{equation}
with:
\begin{eqnarray}\nonumber
&&C (z,t)\equiv\! -\!
\left[\frac{2(u_\mathrm{f}^2\! -\! u_\mathrm{s}^2)(1\!+\! u_\mathrm{A}^2)\mathrm{e}^{i \phi_\mathrm{g} }}{(1\!-\!u_\mathrm{f}^2)(1\!-\!u_\mathrm{s}^2)}-\frac{u_\mathrm{f}(u_\mathrm{f}\! +\! u_\mathrm{A}^2)\mathrm{e}^{i\phi^+_\mathrm{f}}}{1\!-\!u_\mathrm{f}} \right.
\\\nonumber
&&
\left.
-\frac{u_\mathrm{f}(u_\mathrm{f}\! -\! u_\mathrm{A}^2)\mathrm{e}^{i\phi^-_\mathrm{f}}}{1\!+\!u_\mathrm{f}}
\!+\! \frac{u_\mathrm{s}(u_\mathrm{s} \!+\! u_\mathrm{A}^2)\mathrm{e}^{i\phi^+_\mathrm{s}}}{1\!-\!u_\mathrm{s}}
\!+\! \frac{u_\mathrm{s}(u_\mathrm{s} \!-\! u_\mathrm{A}^2)\mathrm{e}^{i\phi^-_\mathrm{s}}}{1\!+\!u_\mathrm{s}}\right],
\end{eqnarray}
and:
\begin{eqnarray}\nonumber
&&D (z,t) \equiv -u_\mathrm{f} u_\mathrm{s}
\left\{\frac{2(u_\mathrm{f}^2\! -\! u_\mathrm{s}^2)\frac{1+ u_{\mathrm{A}\|}^2}{u_{\mathrm{A}\|}^2}
\mathrm{e}^{i \phi_\mathrm{g}}}{(1-u_\mathrm{f}^2)(1-u_\mathrm{s}^2)}\right.
\\\nonumber
&&-\left[\frac{1\!-\!u_\mathrm{A}^2}{u_{\mathrm{A}\|}^2} \!+\! \frac{1\!-\!u_\mathrm{f}}{u_\mathrm{f}}\right]
\frac{\mathrm{e}^{i\phi^+_\mathrm{f}}}{1\!-\!u_\mathrm{f}}  
\!-\!\left[\frac{1\!-\!u_\mathrm{A}^2}{u_{\mathrm{A}\|}^2}\! -\! \frac{1\!-\!u_\mathrm{f}}{u_\mathrm{f}}\right]
\frac{\mathrm{e}^{i\phi^-_\mathrm{f}}}{1\!+\!u_\mathrm{f}}
\\\nonumber
&&\left.
+\left[\frac{1\!-\!u_\mathrm{A}^2}{u_{\mathrm{A}\|}^2} \!+\! \frac{1\!-\!u_\mathrm{s}}{u_\mathrm{s}}\right]
\frac{\mathrm{e}^{i\phi^+_\mathrm{s}}}{1\!-\!u_\mathrm{s}}  
\!+\!\left[\frac{1\!-\!u_\mathrm{A}^2}{u_{\mathrm{A}\|}^2}\! -\! \frac{1\!-\!u_\mathrm{s}}{u_\mathrm{s}}\right]
\frac{\mathrm{e}^{i\phi^-_\mathrm{s}}}{1\!+\!u_\mathrm{s}}\right\}.
\end{eqnarray}
Note that in the relativistic limit ($u_\mathrm{s}\!\downarrow\! 0$) $D$ vanishes.
From $\bm{E} = - \bm{v}^1 \times \bm{B}^0$ we have:
\[
E_y^1 (z,t) = B^0_x v_z^1 (z,t) - B^0_z v_x^1 (z,t).
\]
The current density is most easily found as:
\[
j_y^1 (z,t) = - \frac{i \omega}{B^0 \cos\theta} \frac{\gamma p^0}{c_\mathrm{s}^2} v_x^1 (z,t),
\]
and the pressure is derived from \eqref{eq::energy}:
\begin{eqnarray}\nonumber
&&\frac{p^1}{\gamma p^0} \!=\frac{h_+}{4} \frac{u_{\mathrm{A}\bot}^2u_\mathrm{s}}{u_\mathrm{f}^2\! -\! u_\mathrm{s}^2}
\left[\frac{1\!+\!u_\mathrm{s}}{1\!-\!u_\mathrm{s}} \frac{\mathrm{e}^{i\phi_\mathrm{s}^+}}{u_\mathrm{s}}
+\frac{1\!-\!u_\mathrm{s}}{1\!+\!u_\mathrm{s}} \frac{\mathrm{e}^{i\phi_\mathrm{s}^-}}{u_\mathrm{s}}
-\frac{4\mathrm{e}^{i\phi_\mathrm{g}}}{1-u_\mathrm{s}^2}  \right]
 \\\nonumber
&&- \frac{h_+}{4} \frac{u_{\mathrm{A}\bot}^2u_\mathrm{f}}{u_\mathrm{f}^2\! -\! u_\mathrm{s}^2} \left[
\frac{1\!+\!u_\mathrm{f}}{1\!-\!u_\mathrm{f}} \frac{\mathrm{e}^{i\phi_\mathrm{f}^+}}{u_\mathrm{f}}
+\frac{1\!-\!u_\mathrm{f}}{1\!+\!u_\mathrm{f}} \frac{\mathrm{e}^{i\phi_\mathrm{f}^-}}{u_\mathrm{f}}
-\frac{4 \mathrm{e}^{i\phi_\mathrm{g}}}{1-u_\mathrm{f}^2} 
\right]
\end{eqnarray}
which readily leads to the energy density:
\[
\mu^1 (z,t) = \frac{p^1 (z,t)}{c_\mathrm{s}^2}.
\]
In the limit of a Poynting flux dominated wind where $u_\mathrm{f}\simeq u_\mathrm{A} \uparrow 1$ and $p_0\downarrow0$ so $u_\mathrm{s} \simeq c_\mathrm{s} \downarrow 0$ the fast mode can interact coherently with the {\sc gw} and we find growing amplitudes for all components:
\begin{eqnarray}\nonumber
\frac{B_x^1 (z,t)}{B_0} &=& 
\frac{v_z^1 (z,t)}{\sin\theta} = -\frac{v_x^1 (z,t)}{\cos\theta} = 
\frac{\mu^1 (z,t)}{\mu^0\sin\theta}  =
 \\\label{eq::mswsols}
-\frac{E_y^1 (z,t)}{B_0}&\simeq & \frac{h_+}{2} \sin\theta\ \omega z\  \Im\left[\mathrm{e}^{i\phi_\mathrm{g}} \right],\\\nonumber
\frac{B^0_x j_y (z,t)}{\mu^0 \omega} &\simeq& \frac{h_+}{2} \sin^2\theta\  \omega z\  \Re\left[\mathrm{e}^{i \phi_\mathrm{g}} \right].
\end{eqnarray}

\subsection{All ultra-relativistic Alfv{\'e}n wave components}\label{app::allalfvenrel}
First we evaluate $B_y^\prime$ in terms of laboratory quantities (boosted phase velocities etc.~\footnote{We will omit the superscript on first order quantities to avoid confusion with the primes.}):
\begin{eqnarray}\nonumber
&&B_y^\prime = \frac{h_\times \gamma_{u_\mathrm{A}}^2 B^0_x}{4\gamma}  
\left\{4
 [(1+\beta^2)(1+u_{\mathrm{A}\|}^2) - 4 \beta u_{\mathrm{A}\|} ] \mathrm{e}^{i\phi_\mathrm{g}}
 \right. \\*\label{eq::relBy}
&&\left.
\haak{\frac{1+u_{\mathrm{A}\|}}{1+\beta}}^2
\frac{\mathrm{e}^{i \phi^+_\mathrm{A}}}{\gamma^4}
+ \haak{\frac{1+\beta}{1+u_{\mathrm{A}\|}}}^2
\frac{\mathrm{e}^{i\phi^-_\mathrm{A}}}{\gamma_{u_\mathrm{A}}^4}
\right\}.
\end{eqnarray}
Now we can express all other components of the Alfv{\'e}n wave in terms of \eqref{eq::relBy} and \eqref{eq::relvy}:
\begin{eqnarray}\nonumber
B_y &=& \gamma(B_y^\prime + \beta E_x^\prime) = \gamma B_y^\prime - \beta \gamma B^0_x v_y, \\\nonumber
E_x &=& \gamma(E_x^\prime + \beta B_y^\prime) = \gamma \beta B_y^\prime -  \gamma B^0_x v_y, \\\nonumber
E_z &=& E_z^\prime = - B_x^\prime v_y^\prime = - B_x v_y,\\\nonumber
j_x &=& j_x^\prime = \frac{i \omega}{8 \pi \gamma^2 \gamma_{u_\mathrm{A}}^2} \frac{B^0_z v_y}{(u_\mathrm{A} - \beta)^2},\\\nonumber
j_z &=& \gamma(j_z^\prime +\beta\tau^\prime) = \frac{B_x^0}{8\pi} \left[i\omega v_y \!-\! 2\gamma\beta \frac{\partial v_y}{\partial z} \right] ,\\\nonumber
\tau &=& \gamma(\tau^\prime +\beta j_z^\prime) = \frac{B_x^0}{8\pi} \left[i\omega \beta v_y \!-\! 2\gamma \frac{\partial v_y}{\partial z} \right].
\end{eqnarray}

\subsection{All ultra-relativistic MSW components}\label{app::allmswrel}
We will only consider the limit of a Poynting flux dominated ultra-relativistic wind and Lorentz transform \eqref{eq::mswsols} (with $\omega = \gamma(\omega^\prime + \beta k^\prime) \simeq 2 \gamma \omega^\prime$):
\begin{eqnarray}\nonumber
B_x &=& \gamma(B_x^\prime - \beta E_y^\prime) \simeq 2 \gamma B_x^\prime = 
\frac{h_+}{2 \gamma^2} B_x^0 \omega z\  \Im\left[\mathrm{e}^{i\phi_\mathrm{g}} \right],\\\nonumber
E_y &=& \gamma(E_y^\prime - \beta B_x^\prime) \simeq - B_x,\\\nonumber
j_y &=& j_y^\prime = \frac{h_+}{8\gamma^3} \frac{\mu^0}{B_x^0} \frac{\sin^2\theta}{(1-\beta\cos\theta)^2}\omega^2 z\  \Re\left[\mathrm{e}^{i\phi_\mathrm{g}} \right],\\\nonumber
\mu &=& \gamma \mu^\prime = \frac{h_+}{4 \gamma^2} \frac{\mu^0 \sin\theta}{1-\beta\cos\theta} \omega z\  \Im\left[\mathrm{e}^{i\phi_\mathrm{g}} \right].
\end{eqnarray}

\bibliography{prd} 

\begin{thebibliography}{18}
\expandafter\ifx\csname natexlab\endcsname\relax\def\natexlab#1{#1}\fi
\expandafter\ifx\csname bibnamefont\endcsname\relax
  \def\bibnamefont#1{#1}\fi
\expandafter\ifx\csname bibfnamefont\endcsname\relax
  \def\bibfnamefont#1{#1}\fi
\expandafter\ifx\csname citenamefont\endcsname\relax
  \def\citenamefont#1{#1}\fi
\expandafter\ifx\csname url\endcsname\relax
  \def\url#1{\texttt{#1}}\fi
\expandafter\ifx\csname urlprefix\endcsname\relax\def\urlprefix{URL }\fi
\providecommand{\bibinfo}[2]{#2}
\providecommand{\eprint}[2][]{\url{#2}}

\bibitem[{\citenamefont{{Gertsenshtein}}(1961)}]{gertsenshtein}
\bibinfo{author}{\bibfnamefont{M.}~\bibnamefont{{Gertsenshtein}}},
  \bibinfo{journal}{{Zh. Exsp. Teor. Fiz.}} \textbf{\bibinfo{volume}{{41}}},
  \bibinfo{pages}{113} (\bibinfo{year}{1961}).

\bibitem[{\citenamefont{{Gayer} and {Kennel}}(1979)}]{gayer}
\bibinfo{author}{\bibfnamefont{S.}~\bibnamefont{{Gayer}}} \bibnamefont{and}
  \bibinfo{author}{\bibfnamefont{C.~F.} \bibnamefont{{Kennel}}},
  \bibinfo{journal}{\prd} \textbf{\bibinfo{volume}{19}}, \bibinfo{pages}{1070}
  (\bibinfo{year}{1979}).

\bibitem[{\citenamefont{{Thorne}}(1989)}]{kip}
\bibinfo{author}{\bibfnamefont{K.~S.} \bibnamefont{{Thorne}}}
  (\bibinfo{year}{1989}), \bibinfo{note}{\\{\tt
  http://elmer.tapir.caltech.edu/ph237/}}.

\bibitem[{\citenamefont{{Macedo} and {Nelson}}(1983)}]{macedo}
\bibinfo{author}{\bibfnamefont{P.~G.} \bibnamefont{{Macedo}}} \bibnamefont{and}
  \bibinfo{author}{\bibfnamefont{A.~H.} \bibnamefont{{Nelson}}},
  \bibinfo{journal}{Phys.~Rev.~D.} \textbf{\bibinfo{volume}{28}},
  \bibinfo{pages}{2382} (\bibinfo{year}{1983}).

\bibitem[{\citenamefont{{Moortgat} and {Kuijpers}}(2003)}]{moortgat}
\bibinfo{author}{\bibfnamefont{J.}~\bibnamefont{{Moortgat}}} \bibnamefont{and}
  \bibinfo{author}{\bibfnamefont{J.}~\bibnamefont{{Kuijpers}}},
  \bibinfo{journal}{\aap} \textbf{\bibinfo{volume}{402}}, \bibinfo{pages}{905}
  (\bibinfo{year}{2003}).

\bibitem[{\citenamefont{{Servin}}(2003)}]{servinthesis}
\bibinfo{author}{\bibfnamefont{M.}~\bibnamefont{{Servin}}}, Ph.D. thesis,
  \bibinfo{school}{Department of Physics, Ume{\aa} University, Sweden}
  (\bibinfo{year}{2003}).

\bibitem[{\citenamefont{{Papadopoulos}
  et~al.}(2001)\citenamefont{{Papadopoulos}, {Stergioulas}, {Vlahos}, and
  {Kuijpers}}}]{papadopoulos2}
\bibinfo{author}{\bibfnamefont{D.}~\bibnamefont{{Papadopoulos}}},
  \bibinfo{author}{\bibfnamefont{N.}~\bibnamefont{{Stergioulas}}},
  \bibinfo{author}{\bibfnamefont{L.}~\bibnamefont{{Vlahos}}}, \bibnamefont{and}
  \bibinfo{author}{\bibfnamefont{J.}~\bibnamefont{{Kuijpers}}},
  \bibinfo{journal}{A\&A} \textbf{\bibinfo{volume}{377}}, \bibinfo{pages}{701}
  (\bibinfo{year}{2001}).

\bibitem[{\citenamefont{{Misner} et~al.}(1973)\citenamefont{{Misner}, {Thorne},
  and {Wheeler}}}]{gravitation}
\bibinfo{author}{\bibfnamefont{C.~W.} \bibnamefont{{Misner}}},
  \bibinfo{author}{\bibfnamefont{K.~S.} \bibnamefont{{Thorne}}},
  \bibnamefont{and} \bibinfo{author}{\bibfnamefont{J.~A.}
  \bibnamefont{{Wheeler}}}, \emph{\bibinfo{title}{{Gravitation}}}
  (\bibinfo{publisher}{San Francisco: W.H.~Freeman and Co., 1973},
  \bibinfo{year}{1973}).

\bibitem[{\citenamefont{{Zel{'}dovich}}(1973)}]{zeldovich}
\bibinfo{author}{\bibfnamefont{Y.~B.} \bibnamefont{{Zel{'}dovich}}},
  \bibinfo{journal}{{Zh. Eksp. Teor. Fiz.}} \textbf{\bibinfo{volume}{{65}}},
  \bibinfo{pages}{1311} (\bibinfo{year}{1973}).

\bibitem[{\citenamefont{{Papadopoulos}}(2002)}]{papadopoulosnonlinear}
\bibinfo{author}{\bibfnamefont{D.}~\bibnamefont{{Papadopoulos}}},
  \bibinfo{journal}{\aap} \textbf{\bibinfo{volume}{396}}, \bibinfo{pages}{1045}
  (\bibinfo{year}{2002}).

\bibitem[{\citenamefont{{Servin} and {Brodin}}(2003)}]{servin6}
\bibinfo{author}{\bibfnamefont{M.}~\bibnamefont{{Servin}}} \bibnamefont{and}
  \bibinfo{author}{\bibfnamefont{G.}~\bibnamefont{{Brodin}}},
  \bibinfo{journal}{\prd} \textbf{\bibinfo{volume}{68}}, \bibinfo{pages}{44017}
  (\bibinfo{year}{2003}).

\bibitem[{\citenamefont{{Ellis} and {van Elst}}(1999)}]{ellisbook}
\bibinfo{author}{\bibfnamefont{G.~F.~R.} \bibnamefont{{Ellis}}}
  \bibnamefont{and} \bibinfo{author}{\bibfnamefont{H.}~\bibnamefont{{van
  Elst}}}, in \emph{\bibinfo{booktitle}{Theoretical and Observational
  Cosmology}}, edited by
  \bibinfo{editor}{\bibfnamefont{M.}~\bibnamefont{{Lachi\`{e}ze-Rey}}}
  (\bibinfo{publisher}{Kluwer}, \bibinfo{address}{Dordrecht},
  \bibinfo{year}{1999}), pp. \bibinfo{pages}{p. 1--116}.

\bibitem[{\citenamefont{{Thorne} and {MacDonald}}(1982)}]{mcdonald}
\bibinfo{author}{\bibfnamefont{K.~S.} \bibnamefont{{Thorne}}} \bibnamefont{and}
  \bibinfo{author}{\bibfnamefont{D.}~\bibnamefont{{MacDonald}}},
  \bibinfo{journal}{MNRAS} \textbf{\bibinfo{volume}{198}}, \bibinfo{pages}{339}
  (\bibinfo{year}{1982}).

\bibitem[{\citenamefont{{Brodin} and {Marklund}}(1999)}]{servin1}
\bibinfo{author}{\bibfnamefont{G.}~\bibnamefont{{Brodin}}} \bibnamefont{and}
  \bibinfo{author}{\bibfnamefont{M.}~\bibnamefont{{Marklund}}},
  \bibinfo{journal}{Phys. Rev. Lett.} \textbf{\bibinfo{volume}{82}},
  \bibinfo{pages}{3012} (\bibinfo{year}{1999}).

\bibitem[{\citenamefont{{Achterberg}}(1983)}]{bram}
\bibinfo{author}{\bibfnamefont{A.}~\bibnamefont{{Achterberg}}},
  \bibinfo{journal}{\pra} \textbf{\bibinfo{volume}{28}}, \bibinfo{pages}{2449}
  (\bibinfo{year}{1983}).

\bibitem[{\citenamefont{{Melrose}}(1986)}]{melrose}
\bibinfo{author}{\bibfnamefont{D.~B.} \bibnamefont{{Melrose}}},
  \emph{\bibinfo{title}{{Instabilities in space and laboratory plasmas}}}
  (\bibinfo{publisher}{Cambridge and New York, Cambridge University Press,
  1986, 290 p.}, \bibinfo{year}{1986}).

\bibitem[{\citenamefont{{Brodin} et~al.}(2001)\citenamefont{{Brodin},
  {Marklund}, and {Servin}}}]{photonfreq}
\bibinfo{author}{\bibfnamefont{G.}~\bibnamefont{{Brodin}}},
  \bibinfo{author}{\bibfnamefont{M.}~\bibnamefont{{Marklund}}},
  \bibnamefont{and} \bibinfo{author}{\bibfnamefont{M.}~\bibnamefont{{Servin}}},
  \bibinfo{journal}{\prd} \textbf{\bibinfo{volume}{63}},
  \bibinfo{pages}{124003} (\bibinfo{year}{2001}).

\bibitem[{\citenamefont{{Ibrahim} et~al.}(2003)\citenamefont{{Ibrahim},
  {Swank}, and {Parke}}}]{ibrahim}
\bibinfo{author}{\bibfnamefont{A.~I.} \bibnamefont{{Ibrahim}}},
  \bibinfo{author}{\bibfnamefont{J.~H.} \bibnamefont{{Swank}}},
  \bibnamefont{and} \bibinfo{author}{\bibfnamefont{W.}~\bibnamefont{{Parke}}},
  \bibinfo{journal}{\apjl} \textbf{\bibinfo{volume}{584}}, \bibinfo{pages}{L17}
  (\bibinfo{year}{2003}).

\end{thebibliography}

\end{document}